\documentclass{emulateapj}

\usepackage{natbib}
\usepackage{graphicx}

\setcitestyle{aysep={},citesep={,}}

\newcommand{\eg}{e.g.,}
\newcommand{\ie}{i.e.,}
\newcommand{\magarc}{mag arcsec$^{-2}$}

\newcommand{\fig}[1]{Figure~\ref{fig:#1}}
\newcommand{\sect}[1]{\S\ref{sect:#1}}
\newcommand{\tab}[1]{Table~\ref{tab:#1}}

\newcommand{\zsim}[1]{\ensuremath{z\!\simeq\!{#1}}}
\newcommand{\zeq}[1]{\ensuremath{z\!=\!{#1}}}
\newcommand{\LSun}{\ensuremath{L_{\odot}}}
\newcommand{\MSun}{\ensuremath{M_{\odot}}}
\newcommand{\degree}{\ensuremath{^{\circ}}}
\newcommand{\color}[2]{\ensuremath{(#1 \!-\! #2)}}
\newcommand{\ee}[2]{\ensuremath{#1\times10^{#2}}}

\newcommand{\galfit}{\texttt{GalFit}}
\newcommand{\tinytim}{\texttt{TinyTim}}
\newcommand{\multidrizzle}{\texttt{Multidrizzle}}

\begin{document}
	
\title{Near-Infrared Imaging of a \zeq{6.42} Quasar Host Galaxy With the Hubble Space Telescope Wide Field Camera 3\footnote{Based on observations made with the NASA/ESA Hubble Space Telescope, which is operated by the Association of Universities for Research in Astronomy, Inc., under NASA contract NAS 5-26555. These observations are associated with Program ID 12332.}}

\author{M. Mechtley\altaffilmark{1}, R. A. Windhorst\altaffilmark{1,2}, R. E. Ryan\altaffilmark{3}, G. Schneider\altaffilmark{4}, S. H. Cohen\altaffilmark{1,2}, R. A. Jansen\altaffilmark{1,2}, X. Fan\altaffilmark{4}, N. P. Hathi\altaffilmark{5}, W. C. Keel\altaffilmark{6}, A. M. Koekemoer\altaffilmark{3}, H. R\"ottgering\altaffilmark{7}, E. Scannapieco\altaffilmark{1}, D. P. Schneider\altaffilmark{8,9}, M. A. Strauss\altaffilmark{10}, H. J. Yan\altaffilmark{11}}

\altaffiltext{1}{School of Earth and Space Exploration, Arizona State University, P.O. Box 871404, Tempe, AZ 85287, USA}
\altaffiltext{2}{Department of Physics, Arizona State University, P.O. Box 871504, Tempe, AZ 85287, USA}
\altaffiltext{3}{Space Telescope Science Institute, Baltimore, MD 21218, USA}
\altaffiltext{4}{Steward Observatory, The University of Arizona, Tucson, AZ 85721, USA}
\altaffiltext{5}{Carnegie Observatories, 813 Santa Barbara Street, Pasadena, CA 91101, USA}
\altaffiltext{6}{Department of Physics and Astronomy, University of Alabama, Box 870324, Tuscaloosa, AL 35487, USA}
\altaffiltext{7}{Leiden Observatory, Leiden University, PO Box 9513, 2300 RA Leiden, The Netherlands}
\altaffiltext{8}{Department of Astronomy and Astrophysics, The Pennsylvania State University, 525 Davey Laboratory, University Park, PA 16802, USA}
\altaffiltext{9}{Institute for Gravitation and The Cosmos, The Pennsylvania State University, University Park, PA 16802}
\altaffiltext{10}{Princeton University Observatory, Princeton, NJ 08544, USA}
\altaffiltext{11}{Department of Physics and Astronomy, The University of Missouri, 701 South College Ave, Columbia, MO 65211, USA}

\shortauthors{Mechtley et al.}
\shorttitle{WFC3 Imaging of a \zeq{6.42} Quasar Host Galaxy}

\begin{abstract}
We report on deep near-infrared F125W (J) and F160W (H) Hubble Space Telescope Wide Field Camera 3 images of the \zeq{6.42} quasar J1148+5251 to attempt to detect rest-frame near-ultraviolet emission from the host galaxy. These observations included contemporaneous observations of a nearby star of similar near-infrared colors to measure temporal variations in the telescope and instrument point spread function (PSF). We subtract the quasar point source using both this direct PSF and a model PSF. 

Using direct subtraction, we measure an upper limit for the quasar host galaxy of $m_J>22.8$, $m_H>23.0$~AB~mag ($2\,\sigma$). After subtracting our best model PSF, we measure a limiting surface brightness from $0\farcs3\!-\!0\farcs5$ radius of $\mu_J > 23.5$, $\mu_H > 23.7$~AB~{\magarc} ($2\,\sigma$). We test the ability of the model subtraction method to recover the host galaxy flux by simulating host galaxies with varying integrated magnitude, effective radius, and S\'ersic index, and conducting the same analysis. These models indicate that the surface brightness limit ($\mu_J > 23.5$~AB~{\magarc}) corresponds to an integrated upper limit of $m_J>22\!-\!23$~AB~mag, consistent with the direct subtraction method. Combined with existing far-infrared observations, this gives an infrared excess $\log(IRX)>1.0$ and corresponding ultraviolet spectral slope $\beta>-1.2\pm0.2$. These values match those of most local luminous infrared galaxies, but are redder than those of almost all local star-forming galaxies and \zsim{6} Lyman break galaxies.
\end{abstract}

\maketitle

\section{Introduction} \label{sect:intro}

Since their discovery by \citet{schmidt63}, quasars have been the best and most easily observed beacons to probe the distant universe. The unification model for active galactic nuclei \citep[AGN, \eg][and references therein]{antonucci93} views quasars as unobscured active nuclei, with a super-massive black hole (SMBH) as the central power-house behind the dominant nonstellar continuum.

Ground-based studies \citep[\eg][]{boroson82, mcleod94} are limited by atmospheric seeing, and show that observing the underlying stellar populations is not trivial. \citet{targett12} have recently imaged \zsim{4} quasar host galaxies using adaptive optics with the European Southern Observatory Very Large Telescope, but these required the best seeing conditions and targets with nearby ($<40$ arcsec) bright stars. Studies with the Hubble Space Telescope (HST) have concentrated on quasars at \zsim{0.5\!-\!3}, yielding important constraints on the morphology, luminosity, and stellar populations of quasar hosts \citep[\eg][]{disney95, mcleod95, bahcall97, mclure99, ridgway01, hutchings02, dunlop03, peng06, zakamska06}.
\notetoeditor{Additional references have been omitted from the above lists due to ApJL reference number limits. Ground: taylor96, willott05, HST: keeton00, mcleod01, kukula01, percival01, floyd04}

\citet{goto09} and \citet{willott11} used ground-based telescopes to detect extended Lyman-$\alpha$ emission out to 15~kpc around the \zeq{6.42} quasar CFHQS J232908-030158. This underscores the need to observe at longer, rest-frame near-ultraviolet wavelengths if emission from the host galaxy is to be differentiated from surrounding gas.

Imaging quasar host galaxies allows one to examine how mergers, starbursts, and other changes in environment affect the central nucleus. Observations of molecular gas and the far-infrared continua of \zsim{6} quasars have suggested extremely high average star formation rates of $\gtrsim\!1000$~{\MSun}~yr$^{-1}$ \citep{wang10}, implying potentially luminous hosts. As many quasars are hosted in massive galaxies, imaging \zsim{6} quasar hosts may reveal the properties of the most massive first galaxies. The enormous SMBH mass ($>\!10^9$~{\MSun}) associated with Sloan Digital Sky Survey \citep[SDSS,][]{sdss} quasars at \zsim{6} represents a great theoretical challenge that heavily constrains possible formation methods and merger histories \citep{volonteri06, li07}.

SDSS J114816.64+525150.3 (hereafter J1148+5251), is the best-studied member of the \zsim{6} quasar population, having been extensively observed at multiple wavelengths since its discovery by \citet{fan03}. Near-infrared spectroscopy by \citet{willott03} and \citet{iwamuro04} measured the \ion{Mg}{2} and \ion{Fe}{2} features, estimating a mass of \ee{3}{9}~{\MSun} for the SMBH, an accretion rate near the Eddington limit, and an \ion{Fe}{2}/\ion{Mg}{2} ratio consistent with quasars at lower redshifts. Radio observations of CO lines \citep[\eg][]{walter03, riechers09} indicate the presence of \ee{2.2\!-\!2.4}{10}~{\MSun} of high-excitation molecular gas extending to $\simeq\!2.5$~kpc ($r=0\farcs45$). Studies of the [\ion{C}{2}] line at $158\mu m$ \citep{maiolino05,walter09} provide evidence that the quasar host galaxy is undergoing a vigorous starburst, with an estimated star formation rate density of $\simeq\!1000~\MSun$~yr$^{-1}$~kpc$^{-2}$ extending over kiloparsec scales. Studies of the continuum emission at far-infrared (FIR) wavelengths indicate a warm dust component with an AGN-corrected FIR luminosity of \ee{9.2}{12}~{\LSun} \citep{wang10} and corresponding dust mass of \ee{4.2\!-\!7.0}{8}~{\MSun} \citep{bertoldi03a, robson04, beelen06}. Locally, this is in the range of ultra-luminous infrared galaxies (ULIRGs, galaxies with $L_{FIR}>10^{12}$~{\LSun}). Near- and mid-infrared Spitzer Space Telescope observations by \citet{jiang06} show clear evidence for prominent hot dust within the galaxy. All of these observations argue for a host galaxy with a significant stellar mass component undergoing an extreme episode of star formation.
\notetoeditor{The following additional papers were published on similar data to those listed, again omitted to fall within the reference limit: barth03, bertoldi03b, carilli04, leipski10, walter04}


In this Letter, we report on near-infrared imaging of J1148+5251 with the HST Wide Field Camera 3 (WFC3) and our methods for characterizing and subtracting the instrument and telescope Point Spread Function (PSF). In \sect{obs}, we describe our near-infrared WFC3 imaging of J1148+5251 and a nearby star, used for PSF characterization. In \sect{psfsubtraction}, we describe our method for subtracting the central point source from the quasar images. We describe our simulations for assessing the reliability of our subtraction method in \sect{sims}. Finally, in \sect{discussion}, we discuss the implications of these results and our plans for future investigation. For this paper, we adopt a $\Lambda$CDM cosmology with $H_0=70.3$~km~s$^{-1}$~Mpc$^{-1}$, $\Omega_M=0.271$, and $\Omega_\Lambda=0.729$ \citep{komatsu11}. Unless otherwise stated, all magnitudes use the AB system \citep{oke74} and have been corrected for Galactic extinction using the map of \citet{schlegel98}.


\section{Observations and Data Reduction} \label{sect:obs}

The HST observations of J1148+5251 were performed on 2011 January 31 (HST Program ID 12332, PI: R. Windhorst) using the WFC3 IR channel with the F125W (Wide J) and F160W (WFC3 H) filters. Previous programs \citep[\eg][]{hutchings02} found that the quality of empirical quasar point source subtractions was significantly affected by PSF time variability. This variability is mostly due to so-called ``spacecraft breathing'' effects, \ie thermally induced defocus of the HST secondary mirror due to movement of the Optical Telescope Assembly as the telescope goes into and out of Earth shadow \citep[\eg][]{hershey97}. Examining these studies and instrument reports, we identified two primary sources of thermal variation that we expected to affect our observations --- the spacecraft attitude and the orbital day-night cycle.

To minimize thermal variations due to spacecraft attitude, we constrained our PSF star to be within 5\degree\ of the target quasar, and required that the quasar and PSF star observations be collected in contiguous orbits. We further requested (and were granted) that our observations be scheduled immediately following another observation near the same celestial coordinates, thus minimizing thermal equilibration time in our first orbit. We selected the PSF star to match the quasar near-infrared colors ($\color{J}{H}=0.55$~Vega~mag, $\color{H}{K^{\prime}}=0.72$~Vega~mag) as closely as possible, to minimize differences in wavelength-dependent PSF features. Many stars with colors similar to the quasar also have SDSS spectra, since they were targeted by SDSS as high-redshift quasar candidates. We examined these spectra where available, to reject obvious spectroscopic binaries or other contaminants. After checking the resulting candidate stars for HST guide stars, we selected the star 2MASS~J11552259+4937342 \citep[spectral type K7, $m_J=16.272\pm0.079$~Vega~mag, $\color{J}{H}=0.488\pm0.158$~Vega~mag, $\color{H}{Ks}=0.651\pm0.176$~Vega~mag,][]{2mass} as our final PSF target.

\begin{figure*}[ht!]
\centering
\includegraphics[width=\textwidth]{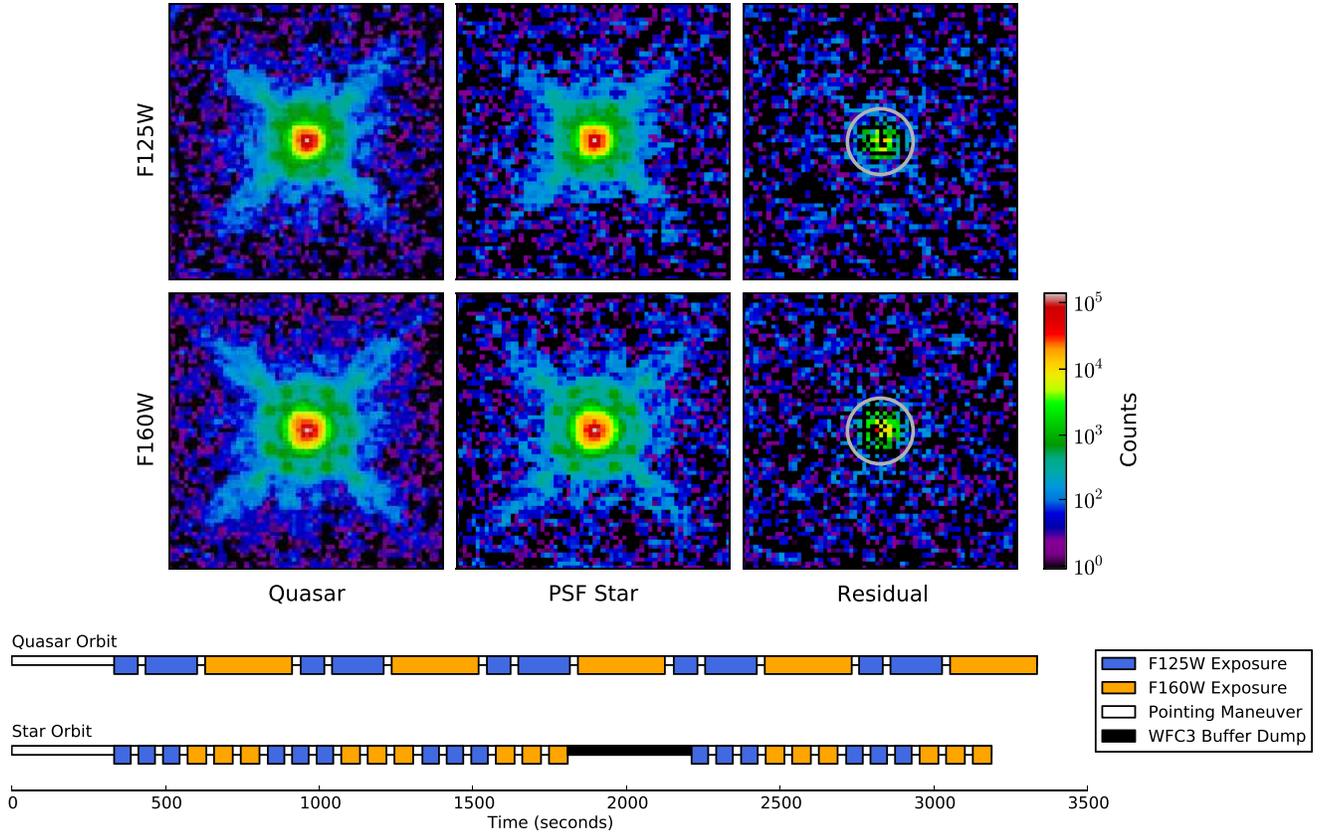}
\caption{Empirical PSF subtraction. Left panels: {\multidrizzle}-combined F125W ($J$, top) and F160W ($H$, bottom) WFC3 images of J1148+5251 after sky subtraction. Pixels are 0\farcs065. Middle Panels: Scaled and shifted PSF star, as fit by {\galfit}. Right panels: Fit residuals, showing a net positive residual flux, but high noise. We measured the integrated residual flux using a 0\farcs5 radius aperture (gray circle), obtaining upper limits of $m_J>22.8$~mag, $m_H>23.0$~mag ($2\,\sigma$). All images are displayed with the same logarithmic stretch.
Bottom panels: PSF star and one quasar orbit, highlighting the relative phasing of corresponding dither points to compensate for spacecraft breathing. Exact phase matches are not possible due to buffer dumps and specific readout sequences. Corresponding dither points were centered at similar positions on the detector to account for field-dependent PSF variability.
}
\label{fig:subtractreal}
\end{figure*}

Unfortunately, there is no way to eliminate thermal variation due to the orbital day/night cycle. Thus, we constructed our orbits to ensure that we fully sampled this cycle, and that equal fractions of the final, combined quasar and PSF star images would come from any given location in orbital phase. Each quasar exposure was matched by several shorter PSF star exposures, taken at the same sub-pixel dither point at the same phase within the orbit. This phase-space sampling is summarized in the bottom panel of \fig{subtractreal}. Any given detector location was never exposed beyond half-well depth in a single orbit, thus avoiding saturation or detector persistence. The quasar was observed in this pattern for three orbits in total, and the PSF star was observed for a single orbit\footnote{An additional PSF star, 2MASS~J11403198+5620582, was also observed for a single orbit to allow for inter-orbit interpolation of the PSF measurement. This observation suffered from a poor guide star acquisition and was unusable.}. The total exposure times for each object and filter combination are summarized in \tab{exptimes}. Analysis was performed on {\multidrizzle}-combined images \citep{multidrizzle,koekemoer11} with an output pixel scale of 0\farcs065, to achieve Nyquist sampling of the PSF core in both filters, enabling accurate spatial shifting. The cosmic ray rejection step of {\multidrizzle} was disabled, since the WFC3 IR \texttt{MULTIACCUM} readout mode provided sufficient cosmic ray rejection.

\begin{deluxetable}{lcrr}[h!]
\tablecolumns{3}
\tablecaption{Summary of Observations}
\tablehead{\colhead{Target} & \colhead{Filter} & \colhead{Exposure} & \colhead{S/N} \\
\colhead{} & \colhead{} & \colhead{Time (s)} & \colhead{}}
\startdata
SDSS J1148+5251 & F125W & 2478 & 2400 \\
SDSS J1148+5251 & F160W & 3646 & 3760 \\
2MASS J11552259+4937342 & F125W & 208 & 1730 \\
2MASS J11552259+4937342 & F160W & 335 & 2200 \\
\enddata
\label{tab:exptimes}
\end{deluxetable}

\section{Point Source Subtraction} \label{sect:psfsubtraction}

To subtract the quasar point source, we used the program {\galfit} \citep{galfit} to fit a PSF single-component model to the quasar image. We transformed the {\multidrizzle}-generated weight maps into uncertainty maps as in \citet{dickinson04}, including the effects of correlated noise and shot noise from the quasar, and supplied these to {\galfit} as the pixel-to-pixel uncertainty (``sigma'') image. We then subtracted the best-fit model from the original image, and inspected the residual.

We first attempted this subtraction using the image of our PSF star as the model PSF. The results are shown in \fig{subtractreal}. We measured the residuals using a 0\farcs5 radius aperture, obtaining upper limits of $m_J>22.8$~mag, $m_H>23.0$~mag ($2\,\sigma$). This includes the total noise contribution from both the quasar and the empirical PSF, measured by scaling the PSF uncertainty map by the same factor as in the fit, and adding it in quadrature to the quasar uncertainty map. The noise contribution from the subtracted PSF is comparable to that of the quasar since the two images have comparable $S/N$ (see \tab{exptimes}), which leads to cosmetic defects (holes) in the subtraction, despite the net positive residual.

\begin{figure*}[ht!]
\centering
\includegraphics[width=0.8\textwidth]{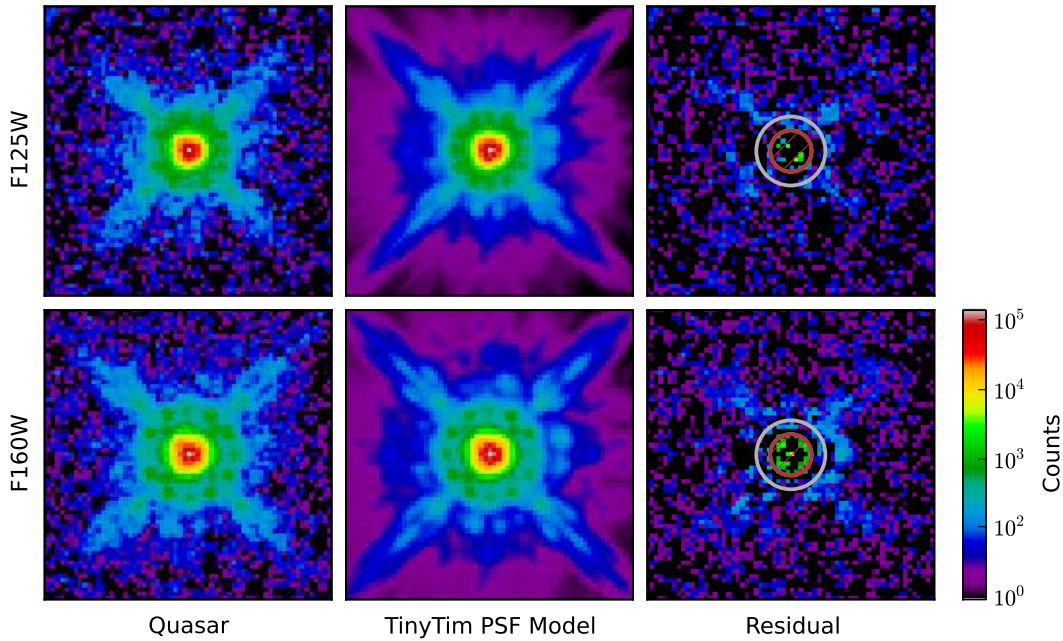}
\caption{Model PSF subtraction. Left panels: {\multidrizzle}-combined F125W ($J$, top) and F160W ($H$, bottom) WFC3 images of J1148+5251. Middle Panels: {\tinytim} models of the quasar point source, constructed by optimizing parameters for the PSF star observations, then scaled and shifted by {\galfit}. Right panels: Fit residuals, showing no significant detection of the underlying galaxy beyond $0\farcs3$ radius. The over-subtracted flux in the central $0\farcs3$ (inner circle) occurs because the best-fit model PSFs have more power in the central peak than the observations, and is also seen in residuals when modeling the PSF star. This region was excluded from the fit. The noise floor in the residual panel is 40\% that of the residual panel in \fig{subtractreal}. From $r=0\farcs3\!-\!0\farcs5$ (between inner and outer circles) we measure a limiting surface brightness of $\mu_J > 23.5$, $\mu_H > 23.7$~{\magarc} ($2\,\sigma$). The noise in the quasar image and uncertainties in the PSF model contribute roughly equally within this region. Visible residuals (diffraction spikes and spots) are also seen when subtracting this model from the PSF star image.
}
\label{fig:subtract}
\end{figure*}

We also generated a {\tinytim}\footnote{http://www.stsci.edu/hst/observatory/focus/TinyTim} \citep{tinytim} model of the PSF, which we calibrated to the the PSF star observations. A $5\times$ spatially oversampled {\tinytim} model was generated for each WFC3 exposure to allow for subpixel shifting. We used the spectrum of J1148+5251 obtained by \citet{iwamuro04} as our model spectrum. We used the HST Focus Model\footnote{http://www.stsci.edu/hst/observatory/focus/FocusModel} \citep{cox11} to estimate the secondary mirror despace for each exposure, and included the field-dependent coma and astigmatism measurements built into {\tinytim}. The individual HST detectors have different mean focus offsets in the Focus Model, but the offset for the WFC3 IR channel has not been characterized. We therefore allowed the mean $Z4$ Zernike coefficient in {\tinytim} \citep[$R^0_2$ in the original formulation of][]{zernike34} to float as a free parameter in our optimization, adding this single mean value to the Focus Model estimate for each exposure. These models for individual exposures were then combined, weighted by exposure time, to produce a composite PSF for each {\multidrizzle}-combined science image. {\galfit} also accepts a pixel response convolution kernel for oversampled PSFs. We generated this quantity in a similar manner, by drizzling copies of the empirical WFC3 pixel response convolution kernel \citep[modeling inter-pixel capacitance and jitter, see][]{hartig08} using the same shifts applied to the real images.

The result of the {\tinytim} PSF subtraction, with a significantly reduced noise floor compared to the direct subtraction, is shown in \fig{subtract}. No host galaxy is detected, to a limiting surface brightness from $r=0\farcs3$ to $0\farcs5$ radius of $\mu_J>23.5$, $\mu_H>23.7$~{\magarc} ($2\,\sigma$). The inner 0\farcs3 was excluded from the fit, as the best-fit {\tinytim} models produce PSF cores that are consistently narrower than those observed. Visible residuals (diffraction spikes and spots) are also seen when subtracting this model from the PSF star observations.

\section{Host Galaxy Simulations} \label{sect:sims}

Having established no host galaxy detection using the {\tinytim} PSF, we sought to quantify this subtraction method's ability to recover the host galaxy flux as a function of host galaxy parameters. To do so, we used {\galfit} to simulate a point source along with a S\'ersic profile host galaxy, with total flux adding up to $m_J=19.1$~mag, the measured flux in F125W. This simulated image contained no noise, so we added both shot noise from the object and a Gaussian noise field drizzled in the same manner as the real quasar image, to match its correlated noise properties. We then performed the same analysis that we used on the real quasar image, using {\galfit} to subtract a {\tinytim}-generated point source and measuring the surface brightness from $r=0\farcs3$ to $0\farcs5$ in the residual image.

We ran a grid of 256 models using this technique, varying the total integrated flux of the host galaxy from $m_J=20\!-\!26$~mag, the effective radius from $r_e=0\farcs1\!-\!0\farcs9$, and S\'ersic indexes $n=1.0$ and $4.0$. The magnitude range represents host galaxies with luminosities from $\simeq1/2$ to $1/500$ of the total quasar luminosity. Fainter host galaxies than this are undetectable due to shot noise from the point source. The range in effective radius corresponds to $r_e\simeq0.6\!-\!5.0$~kpc at \zeq{6.42}.

\begin{figure}[ht!]
\centering
\includegraphics[width=\columnwidth]{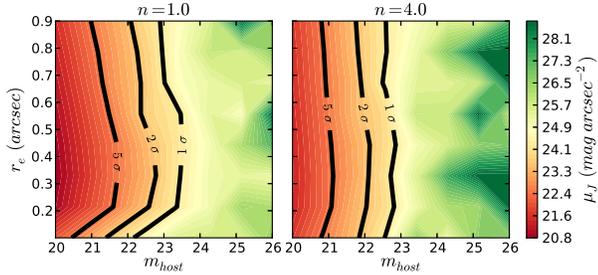}
\caption{PSF subtraction on simulated host galaxies. Surface brightness predicted from $r=0\farcs3\!-\!0\farcs5$ as a function of simulated host galaxy parameters. The simulated observation is approximated as a PSF component along with a S\'ersic profile, with total flux adding up to the observed $m_J$ from our WFC3 image. The integrated magnitude ($m_{host}$), effective radius ($r_e$), and S\'ersic index ($n=1.0$, left panels and $n=4.0$, right panels) of the S\'ersic profile are varied with each model. The $1$, $2$, and $5\,\sigma$ detection significance levels are plotted as black lines. The measured surface brightness reaches the $2\,\sigma$ limit ($\mu_J > 23.5$~{\magarc}) for a host galaxy of $m_J>22-23$~mag, depending upon S\'ersic index and effective radius.
}
\label{fig:sims}
\end{figure}

\fig{sims} summarizes the results of these simulations, plotting the measured surface brightness from $r=0\farcs3\!-\!0\farcs5$ and contours representing the $1$, $2$, and $5\,\sigma$ detection limits. Inspecting the residuals of these model subtractions, we also found that bright ($m_J<22.5$~mag), compact ($r_e<0\farcs3$) host galaxies would cause the method to significantly over-subtract the PSF. This would show negative residuals from the diffraction spikes, which are not seen in \fig{subtract}.

The model surface brightness from $r=0\farcs3\!-\!0\farcs5$ reaches the $2\,\sigma$ upper limit of $\mu_J > 23.5$~{\magarc} for a host galaxy of $m_J>22\!-\!23$~mag, depending upon S\'ersic index and effective radius.

\section{Discussion} \label{sect:discussion}


We have performed point source subtraction on the \zeq{6.42} quasar J1148+5251, with both empirical and modeled PSFs. Using direct subtraction, we measure an upper limit of $m_J>22.8$~mag, $m_H>23.0$~mag ($2\,\sigma$). With the modeled PSF subtraction we measure a limiting surface brightness measured from $0\farcs3\!-\!0\farcs5$ of $\mu_J > 23.5$~{\magarc}, $\mu_H > 23.7$~{\magarc} ($2\,\sigma$). Performing the same subtraction method on simulated quasars, we found that this surface brightness limit corresponds to a host galaxy of $m_J>22\!-\!23$~mag, consistent with the direct subtraction limit.

Using the direct subtraction limits, the upper limits on the rest-frame monochromatic luminosity ($\lambda L_{\lambda}$) at 1700~{\AA} and 2200~{\AA} are $L_{1700}<\ee{8.4}{11}$~{\LSun} and $L_{2200}<\ee{5.4}{11}$~{\LSun}, assuming a flat spectrum within each band when applying the K-correction \citep{oke68}. This is comparable to the most luminous Lyman break galaxies at \zsim{2\!-\!3} \citep{hoopes07}.

Using our upper limits and Equation~1 from \citet{kennicutt98}, which relates $L_\nu$ to star formation rate, we estimate a star formation rate of SFR~$<\!210\!-\!250$~{\MSun}~yr$^{-1}$. This estimate ignores dust attenuation and assumes a continuous star formation rate over $10^8$ years or longer. A younger population would decrease this upper limit, while dust would allow for a higher (absorption-corrected) rate. The star formation rate estimated from the AGN-corrected FIR luminosity by \citet{wang10} is $2380$~{\MSun}~yr$^{-1}$. Since J1148+5251 would be classified as a ULIRG locally, this discrepancy is likely due to significant UV absorption by dust.

We can constrain the infrared excess (IRX) of the host galaxy, defined as the infrared to far-ultraviolet (FUV) luminosity ratio $L_{IR}/L_{FUV}$ \citep[\eg][]{howell10}, usually expressed in logarithmic units. Using our upper limit for $L_{1700}$ and an AGN-corrected infrared luminosity $L_{IR}=\ee{9.2}{12}$~{\LSun} \citep{wang10}, we calculate $\log(IRX)>1.0$, consistent with local luminous infrared galaxies (LIRGs) and ULIRGs \citep{howell10}, but greater than local starburst galaxies and high-redshift Lyman break galaxies \citep{overzier11}.

\begin{figure}[ht!]
\centering
\includegraphics[width=\columnwidth]{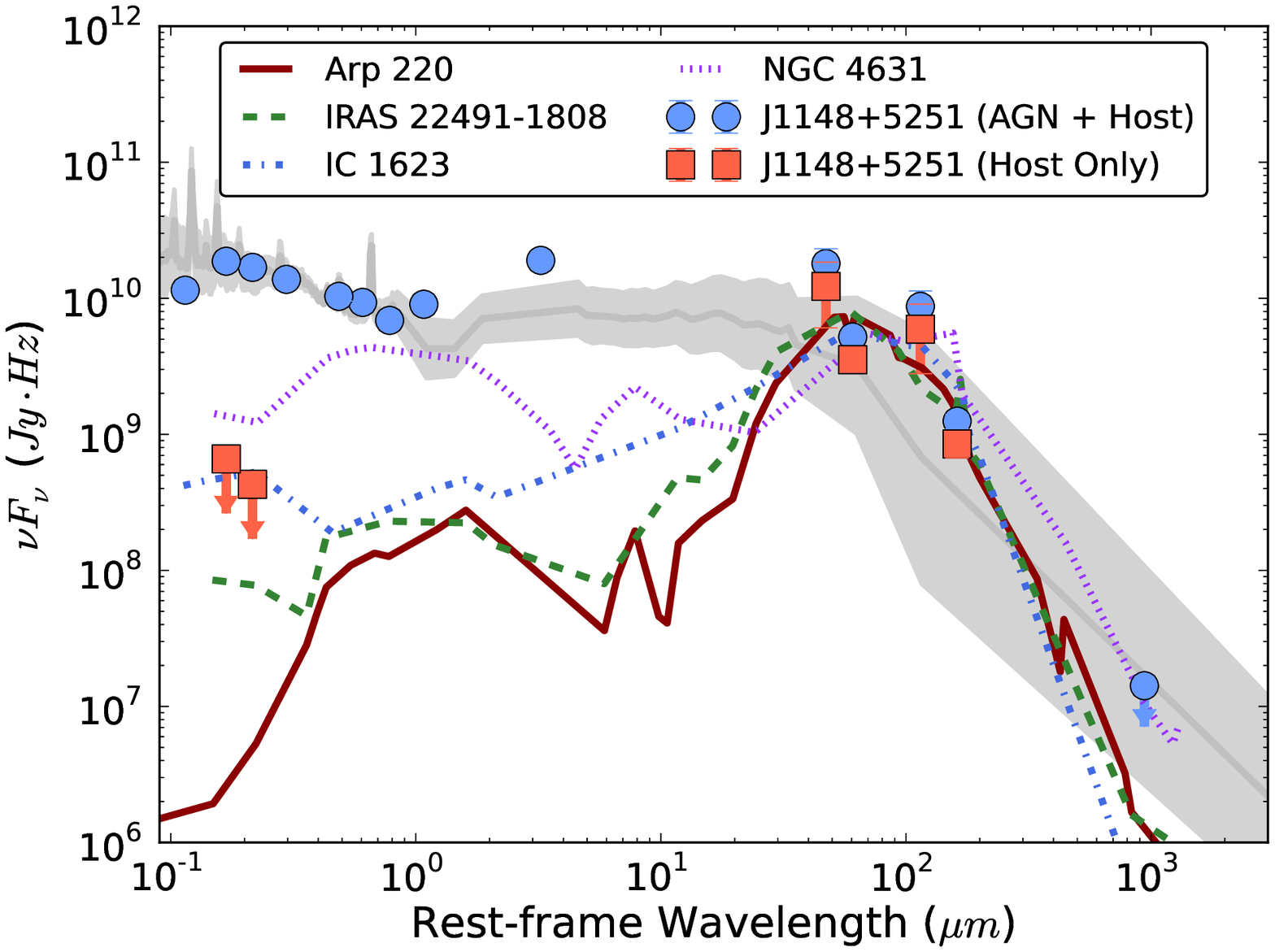}
\caption{Comparison of broad-band photometry for J1148+5251 to local galaxies. Blue circles are broad-band (quasar and host) photometry of J1148+5251 taken from \citet{fan03, iwamuro04, jiang06, beelen06, robson04}, and \citet{bertoldi03a}. Red squares show our upper limits for the host galaxy flux at 1700~{\AA} and 2200~{\AA}, and the AGN-corrected FIR measurements from \citet{wang10}. The light gray spectrum is the average radio-quiet quasar spectrum of \citet{shang11}, normalized to J1148+5251 from $0.1-1 \mu m$. 
The dotted purple SED is NGC~4631, a local spiral with $\log(IRX)<1.0$. Other SEDs are those of the local LIRGs Arp~220 (solid red, $\log(IRX)=3.423$), IRAS~22491-1808 (dashed green, $\log(IRX)=2.198$), and IC~1623 (dot-dashed blue, $\log(IRX)=1.379$), representing high, average, and low IRX LIRGs, respectively \citep{howell10}. The local galaxy SEDs have been normalized to match the AGN-corrected emission of J1148+5251 between rest-frame 40 and 200 microns. Our constraint of $\log(IRX)>1.0$ (and the 2200~{\AA} flux limit) matches most local LIRGs, but is greater than almost all local star-forming galaxies and high-redshift Lyman break galaxies \citep{howell10,overzier11}.
}
\label{fig:sed}
\end{figure}

\fig{sed} plots broad-band measurements for J1148+5251, as well as our upper limits for the host galaxy flux at 1700~{\AA} and 2200~{\AA} and the AGN-corrected FIR measurements of \citet{wang10}. Also plotted are the spectral energy distributions (SEDs) of four local galaxy systems --- three LIRGs (Arp~220, IRAS~22491-1808, and IC~1623), representing the range in IRX from \citet{howell10}, and the star-forming spiral NGC~4631, representing a galaxy with $\log(IRX)<1.0$, which is thus excluded as a potential host. Photometric points for the local galaxies are taken from NED\footnote{The NASA/IPAC Extragalactic Database (NED) is operated by the Jet Propulsion Laboratory, California Institute of Technology, under contract with NASA.} and the SEDs have been normalized to match the AGN-corrected emission of J1148+5251 between 40 and 200 microns.

We estimate $A_{FUV}>2.1$~mag of UV absorption using the relation between the IRX flux ratio and $A_{FUV}$ \citep[\eg][Equation~1]{overzier11}. Using the empirical relation $IRX_{M99,inner}$ ($A_{FUV}=4.54+2.07\beta\pm0.4$) from \citet{overzier11}, we obtain a limit of $\beta>-1.2\pm0.2$. This matches local (U)LIRGs \citep{howell10}, but is redder than almost all local star-forming galaxies and \zsim{6} Lyman break galaxies \citep{overzier11,bouwens12}.

The {\tinytim}-based subtraction may be improved in the future with more accurate WFC3 IR PSFs. Since uncertainties introduced by the PSF model scale with PSF brightness, our further WFC3 observations will target quasars where the contrast ratio between point source and host galaxy is expected to be smaller, such as optically faint \zsim{6} quasars with large FIR luminosities. The James Webb Space Telescope will enable us to use the PSF subtraction method at rest-frame ultraviolet and optical wavelengths with better-sampled empirical PSFs in a more stable thermal environment.

Special thanks are due to K. Long, J. MacKenty, and A. Roman at STScI for their advice and expertise when planning the observations, and the anonymous referee for very useful suggestions. We are grateful to J. Krist and C. Peng for the public availability of {\tinytim} and {\galfit}, respectively. This research has made use of the SIMBAD database, operated at CDS, Strasbourg, France. Support for this HST program was provided by NASA through grant GO-12332.*.A from the Space Telescope Science Institute, which is operated by the Association of Universities for Research in Astronomy, Inc., under NASA contract NAS 5-26555. We owe a deep debt of gratitude to the entire WFC3 instrument team and the astronauts of STS-125 for this marvelous instrument on HST.


\begin{thebibliography}{}

\bibitem[Antonucci(1993)]{antonucci93}Antonucci, R. 1993, \araa, 31, 473
\bibitem[Bahcall et al.(1997)]{bahcall97}Bahcall, J. N., Kirhakos, S., Saxe, D. H., \& Schneider, D. P. 1997, \apj, 479, 642
\bibitem[Beelen et al.(2006)]{beelen06}Beelen, A., Cox, P., Benford, D. J., Dowell, C. D., et al. 2006, \apj, 642, 694
\bibitem[Bertoldi et al.(2003)]{bertoldi03a}Bertoldi, F., Carilli, C. L., Cox, P., Fan, X., et al. 2003, \aap, 406, 55
\bibitem[Boroson \& Oke(1982)]{boroson82}Boroson, T. A. \& Oke, J. B. 1982, \nat, 296, 397
\bibitem[Bouwens et al.(2012)]{bouwens12}Bouwens, R. J., Illingworth, G. D., Oesch, P. A., Franx, M., et al. 2012, arXiv:1109.0994
\bibitem[Cox \& Niemi(2011)]{cox11}Cox, C. \& Niemi, S.-M. 2011, Instrument Science Report, TEL 2011-01, STScI, Baltimore
\bibitem[Dickinson et al.(2004)]{dickinson04}Dickinson, M., Stern, D., Giavalisco, M., Ferguson, H. C., et al. 2004, \apj, 600, L99
\bibitem[Disney et al.(1995)]{disney95}Disney, M. J., et al. 1995, \nat, 376, 150
\bibitem[Dunlop et al.(2003)]{dunlop03}Dunlop, J. S., McLure, R. J., Kukula, M. J., Baum, S. A. et al. 0023, \mnras, 340, 1095
\bibitem[Fan et al.(2003)]{fan03}Fan, X., Strauss, M. A., Schneider, D. P., Becker, R. H., et al. 2003, \aj, 125, 1649
\bibitem[Goto et al.(2009)]{goto09}Goto, T., Utsumi, Y., Furusawa, H., Miyazaki, S., \& Komiyama, Y. 2009, \mnras, 400, 843
\bibitem[Hartig(2008)]{hartig08} Hartig, G. F. 2008, Instrument Science Report, SESD-08-41, STScI, Baltimore
\bibitem[Hershey(1998)]{hershey97} Hershey, J. L. 1997, Instrument Science Report, SESD-97-01, STScI, Baltimore
\bibitem[Hoopes et al.(2007)]{hoopes07}Hoopes, C. G., Heckman, T. M., Salim, S., Seibert, M., et al. 2007, \apjs, 173, 441
\bibitem[Howell et al.(2010)]{howell10}Howell, J. H., Armus, L., Mazzarella, J. M., Evans, A. S., et al. 2010, \apj, 715, 572
\bibitem[Hutchings et al.(2002)]{hutchings02} Hutchings, J. B., Frenette, D., Hanisch, R., Mo, J., et al. 2002, \aj, 123, 2936
\bibitem[Iwamuro et al.(2004)]{iwamuro04}Iwamuro, F., Masahiko, M., Eto, S., Maihara, T., et al. 2004, \apj, 614, 69
\bibitem[Jiang et al.(2006)]{jiang06}Jiang, L., Fan, X., Hines, D. C., Shi, Y., et al. 2006, \aj, 132, 2127
\bibitem[Kennicutt(1998)]{kennicutt98}Kennicutt, R. C., Jr. 1998, \araa, 36, 189
\bibitem[Koekemoer et al.(2002)]{multidrizzle}Koekemoer, A. M., Fruchter, A. S., Hook, R. N. \& Hack, W. 2002 HST Calibration Workshop (ed. S. Arribas, A. Koekemoer, B. Whitmore, Baltimore: STScI), 337
\bibitem[Koekemoer et al.(2011)]{koekemoer11}Koekemoer, A. M., Faber, S. M., Ferguson, H. C., Grogin, N. A. et al. 2011, \apjs, 197, 36
\bibitem[Komatsu et al.(2011)]{komatsu11}Komatsu, E., Smith, K. M., Dunkley, J., Bennett, C. L., et al. 2011, \apjs, 192, 18
\bibitem[Krist, Hook, \& Stoehr(2011)]{tinytim}Krist, J. E., Hook, R. N., \& Stoehr, F. 2011, Proc. of SPIE, 8127, 81270J-1
\bibitem[Li et al.(2007)]{li07}Li, Y., Hernquist, L., Robertson, B., Cox, T. J., et al. 2007, \apj, 665, 187L
\bibitem[Maiolino et al.(2005)]{maiolino05}Maiolino, R., Cox, P., Caselli, P., Beelen, A., et al. 2005, \aap, 440, 51
\bibitem[McLeod \& Rieke(1994)]{mcleod94}McLeod, K. \& Rieke, G. 1994, \apj, 420, 58
\bibitem[McLeod \& Rieke(1995)]{mcleod95}McLeod, K. \& Rieke, G. 1995, \apjl, 454, 77
\bibitem[McLure et al.(1999)]{mclure99}McLure, R. J., Kukula, M. J., Dunlop, J. S., Baum, S. A. et al. 1999, \mnras, 308, 377
\bibitem[Oke(1974)]{oke74}Oke, J. B. 1974, \apjs, 27, 21
\bibitem[Oke \& Sandage(1968)]{oke68}Oke, J. B. \& Sandage, A. 1968, \apj, 154, 210
\bibitem[Overzier et al.(2011)]{overzier11}Overzier, R. A., Heckman, T. M., Wang, J., Armus, L., et al. 2011, \apjl, 726, L7
\bibitem[Peng et al.(2006)]{peng06}Peng, C. Y., Impey, C. D., Rix, H.-W., Kockanek, C. S., et al. 2006, \apj, 649, 616
\bibitem[Peng et al.(2010)]{galfit}Peng, C. Y., Ho, L. C., Impey, C. D., \& Rix, H.-W. 2010, \aj, 139, 2097
\bibitem[Ridgway et al.(2001)]{ridgway01} Ridgway S. E., Heckman T. M., Calzetti D., \& Lehnert M. 2001, \apj, 550, 122
\bibitem[Riechers et al.(2009)]{riechers09}Riechers, D. A., Walter, F., Bertoldi, F., Carilli, C. L., et al. 2009, \apj, 703, 1338
\bibitem[Robson et al.(2004)]{robson04}Robson, I., Priddey, R. S., Isaak, K. G., \& McMahon, R. G. 2004, \mnras, 351, L29
\bibitem[Schlegel et al.(1998)]{schlegel98}Schlegel, D. J., Finkbeiner, D. P., \& Davis, M. 1998, \apj, 500, 525
\bibitem[Schmidt(1963)]{schmidt63}Schmidt, M. 1963, \nat, 197, 1040
\bibitem[Shang et al.(2011)]{shang11}Shang, Z., Brotherton, M. S., Wills, B. J., Wills, D., et al. 2011, \apjs, 196, 2
\bibitem[Skrutskie et al.(2006)]{2mass}Skrutskie, M. F., Cutri, R. M., Stiening, R., Weinberg, M. D., et al. 2006, \aj, 131, 1163
\bibitem[Targett, Dunlop, \& McLure(2012)]{targett12}Targett, T. A., Dunlop, J. S., \& McLure, R. J 2012, \mnras, 420, 3621
\bibitem[Volonteri \& Rees(2006)]{volonteri06}Volonteri, M., \& Rees, M. J. 2006, \apj, 650, 669
\bibitem[Walter et al.(2003)]{walter03}Walter, F., Bertoldi, F., Carilli, C., Cox, P., et al. 2003, \nat, 424, 406
\bibitem[Walter et al.(2009)]{walter09}Walter, F., Riechers, D., Cox, P., Neri, R., et al. 2009, \nat, 457, 699
\bibitem[Wang et al.(2010)]{wang10}Wang, R., Carilli, C. L., Neri, R., Riechers, D. A., et al. 2010, \apj, 714, 699
\bibitem[Willott et al.(2003)]{willott03}Willott, C. J., McLure, R. J., \& Jarvis, M. J. 2003, \apj, 587, 15
\bibitem[Willott et al.(2011)]{willott11}Willott, C. J., Chet, S., Bergeron, J., \& Hutchings, J. B. 2011, \aj, 142, 186
\bibitem[York et al.(2000)]{sdss}York, D. G., Adelman, J., Anderson, J. E., Jr., Anderson, S. F., et al. 2000, \aj, 120, 1579
\bibitem[Zakamska et al.(2006)]{zakamska06}Zakamska, N. L., Strauss, M. A., Krolik, J. H., Ridgway, S. E., et al. 2006, \aj, 132, 1496
\bibitem[Zernike(1934)]{zernike34}Zernike, F. 1934, \mnras, 94, 377
\end{thebibliography}
\end{document}